\documentclass[aps,prd,showpacs,amsmath,amssymb,reprint]{revtex4-1}
\usepackage{hyperref}
\usepackage{graphicx}
\usepackage{color}

\begin{document}

\title{Equivalence between the Weyl-tensor and gauge-invariant graviton two-point functions in Minkowski and de~Sitter spaces}

\author{Atsushi Higuchi}
\email{atsushi.higuchi@york.ac.uk}
  \affiliation{Department of Mathematics, University of York, Heslington,
York YO10 5DD, United Kingdom}

\date{August 21, 2013}

\begin{abstract}
\noindent
\emph{Revised and extended version of a talk given at the Second Workshop on Quantum Field Theory, August 24-27, 2011, in S\~ao Lu\'{\i}s, Brazil, based on a joint work with Christopher J.\ Fewster.}

\smallskip

The two-point Wightman function of the free photon field defined in a gauge-invariant manner is known to be equivalent to the field-strength two-point function in any spacetime that is topologically trivial.  We show that the gauge-invariant graviton two-point function defined in a similar manner is equivalent to the Weyl-tensor two-point function in Minkowski space and in the Poincar\'e patch of de~Sitter space.  This implies that in the Poincar\'e patch of de~Sitter space the gauge-invariant graviton two-point function decays like $({\rm distance})^{-4}$ as a function of coordinate distance for spacelike separation.
%and decays exponentially in the timelike direction.
\end{abstract}

%\pacs{03.70.+k,04.62+v}
\maketitle

\section{Introduction}

The inflationary cosmology~\cite{Guth,Linde,Steinhardt,Sato1,Sato2,Kazanas}, which assumes an era of exponential expansion in very early universe, has been the leading candidate in addressing various conceptual issues in the standard big-bang cosmology such as the flatness and causality problems.  The exponentially expanding universe in inflationary cosmology is approximately the expanding half of de~Sitter space.  For this reason quantum field theory in de~Sitter space has been studied recently by many authors.
In particular, the infrared (IR) properties of the graviton two-point function in this spacetime has been a subject of lively debate over the past
three decades~\cite{FordParker2,HiguchiNP,Allen86a,AntoniadisMottola,HiguchiCQG,Kleppe,Higuchi:2011vw,Miao:2011ng}.

The two-point function in the transverse-traceless-synchronous gauge in the Poincar\'e patch, i.e.\ the spatially-flat expanding half of de~Sitter space, is IR divergent in the sense that it is not well-defined unless the IR cutoff is inserted~\cite{FordParker2}. This fact and other observations have led some authors to claim that there is no de~Sitter-invariant vacuum state for linearized gravity~\cite{Kleppe,Miao:2011ng} in spite of the fact that IR-finite two-point functions have been constructed in other gauges~\cite{AllenTuryn,HKcovariant,HawkingHertogTurok,HiguchiWeeks}.

The graviton two-point function also diverges or tends to a constant as the separation of the two points tends to infinity both in the spacelike and timelike directions~\cite{Allen86a}.  No gauge conditions are known that make the graviton two-point function tend to zero as the separation of the two points becomes infinite.  If linearized gravity were not a gauge theory, this behavior of the two-point function would imply that the graviton field had strong long-distance correlation.  However, since linearized gravity \emph{is} a gauge theory, one needs to characterize the long-distance correlation in a gauge-invariant manner.  For example, some cosmologists argue that the logarithmic growth of the two-point function for graviton as well as for minimally-coupled massless scalar field cannot be observed~\cite{Urakawa1,Tanaka,Urakawa2,Hebecker1,Hebecker2,Seery}.
However, there is no consensus in the scientific community about the physical significance of the long-distance behavior of the graviton two-point function (see, e.g., Refs.~\cite{GarrigaTanaka,TsamisWoodard}).

Recently a gauge-invariant graviton two-point function has been formulated~\cite{FewsterHunt} following a similar construction for electromagnetic field~\cite{Dimock,FewsterPfenning}.  The main purpose of this paper is to show that this graviton two-point function reduces to that of the Weyl tensor in the Poincar\'e patch of de~Sitter space, which decays at large distances.

The rest of this paper is organized as follows.  In Sec.~\ref{gauge-inv} we motivate and introduce the gauge-invariant graviton two-point function. In Secs.~\ref{Minkowski} and \ref{deSitter} we show the equivalence of the Weyl-tensor and gauge-invariant graviton two-point functions in Minkowski and de~Sitter spaces.  Then, we discuss our result in Sec.~\ref{discuss}. In the Appendix we present a technical result used in Sec.~\ref{Minkowski}.

\section{Gauge-invariant graviton two-point function} \label{gauge-inv}

Let us quickly introduce the gauge-invariant graviton two-point function proposed in Ref.~\cite{FewsterHunt} in a heuristic manner.  To this end we first briefly discuss quantization of the free graviton field.  We linearize the gravitational field about a globally-hyperbolic spacetime satisfying the vacuum Einstein equations with or without a cosmological constant.  We write the Lagrangian density for the linearized gravitational field, or the graviton field, $h_{ab}$ as
\begin{equation}
\mathcal{L} = \frac{\sqrt{-g}}{2}\left[ K^{abca'b'c'}\nabla_a h_{bc} \nabla_{a'}h_{b'c'} + S^{aba'b'}h_{ab}h_{a'b'}\right],
\end{equation}
where $K^{abca'b'c'}=K^{a'b'c'abc}=K^{acba'b'c'}$ and
$S^{aba'b'}=S^{a'b'ab}=S^{baa'b'}$.  We define the conjugate momentum current as
\begin{equation}
p^{abc} := K^{abca'b'c'}\nabla_{a'}h_{b'c'}, \label{conj-mom-curr}
\end{equation}
then the Euler-Lagrange field equation for $h_{bc}$ is
\begin{equation}
-\nabla_a p^{abc} + S^{bcb'c'}h_{b'c'} = 0.  \label{E-L}
\end{equation}

For any two solutions, $h^{(n)}_{bc}$ and $h^{(m)}_{bc}$, and their conjugate momentum currents, $p^{(n)abc}$ and $p^{(m)abc}$, we define the symplectic product as
\begin{equation}
(h^{(n)},h^{(m)})_{\rm symp}
:= \int_{\Sigma} d\Sigma n_a (h^{(n)}_{bc}p^{(m)abc} - p^{(n)abc}h^{(m)}_{bc}),
\end{equation}
where $\Sigma$ is any Cauchy surface and where $n^a$ is the past-directed normal to $\Sigma$.  One can readily see that this product is conserved, i.e.\ independent of the Cauchy surface $\Sigma$~\cite{Friedman,WaldZoupas}.  The symplectic product is degenerate because any tensor of the form $h^{(g)}_{ab} = \nabla_a \Lambda_b + \nabla_b \Lambda_a$ is a solution to Eq.~(\ref{E-L}) and
$(h^{(g)},h^{(m)})_{\rm symp}=0$ for all solutions $h^{(m)}_{ab}$. This means that the symplectic product is gauge invariant, i.e.\ if $h^{(n)}_{ab}$ and $h^{(m)}_{ab}$ are two solutions and if
$h^{(n')}_{ab} = h^{(n)}_{ab} + \nabla_a \Lambda_b^{(n)} + \nabla_b \Lambda_a^{(n)}$ for some
$\Lambda_a^{(n)}$ and similarly for $h^{(m')}_{ab}$, then
$(h^{(n)},h^{(m)})_{\rm symp} = (h^{(n')},h^{(m')})_{\rm symp}$.

In the spacetimes we are interested in, it is possible to impose gauge conditions such that the symplectic product is nondegenerate on the space of solutions satisfying them~\cite{FewsterHunt}.
We expand the quantum field $\hat{h}_{ab}(x)$ in terms of a complete set of solutions $h_{ab}^{(n)}(x)$ satisfying these gauge conditions:
%\begin{equation}
$\hat{h}_{ab}(x) = \sum_n \hat{a}_n h_{ab}^{(n)}(x)$.
%\end{equation}
The field $\hat{h}_{ab}(x)$ is quantized through the commutation
relations $[\hat{a}_n,\hat{a}_m] = (\Omega^{-1})_{nm}$, where $\Omega^{-1}$ is the inverse of the matrix $\Omega^{nm} := (h^{(n)},h^{(m)})_{\rm symp}$.

The field operator $\hat{h}_{ab}(x)$ changes under gauge transformations, but the operators $\hat{a}_n$ are invariant since they can be expressed using the gauge-invariant symplectic product as
$\hat{a}_n = (\Omega^{-1})_{nm}(h^{(m)},\hat{h})_{\rm symp}$.
Thus, all gauge-invariant content of the field $\hat{h}_{ab}(x)$ can be extracted
if $(h^{(n)},\hat{h})_{\rm symp}$ are known for all solutions $h_{ab}^{(n)}(x)$.
In fact it is sufficient for this symplectic product to be known for all solutions with compact support on a Cauchy surface $\Sigma$.

If the solution $h^{(n)}_{ab}(x)$ is compactly supported on $\Sigma$, then one can find a transverse (or divergence-free)
tensor $f^{(n)ab}(x)$ compactly supported in spacetime such that
$(h^{(n)},\hat{h})_{\rm symp} = (f^{(n)},\hat{h})_{\rm st}$, where
\begin{equation}
(f^{(n)},\hat{h})_{\rm st} :=
\int d^Dx\,\sqrt{-g(x)}\,f^{(n)ab}(x)\hat{h}_{ab}(x). \label{spacetime-smear}
\end{equation}
(Here, $D$ is the spacetime dimension.)
This can be shown as follows by generalizing the scalar case~\cite{KayWald,Waldtextbook}.  We define $h^{(n,\chi)}_{ab}(x) :=\chi(x)h^{(n)}_{ab}(x)$, where $\chi(x)=1$ in the future of $\Sigma$ and $\chi(x)=0$ in the past of another Cauchy surface $\Sigma'$.  The function $\chi$ changes its value smoothly from $1$ to $0$ between $\Sigma$ and $\Sigma'$. Let
$p^{(n,\chi)abc}$ be the conjugate momentum current of $h^{(n,\chi)}_{ab}$ as given by Eq.~(\ref{conj-mom-curr}) and define
\begin{equation}
f^{(n)bc} := -\nabla_a p^{(n,\chi)abc} + S^{bcb'c'}h^{(n,\chi)}_{b'c'}.
\end{equation}
Since the right-hand side is the linearized Einstein tensor, the tensor $f^{(n)bc}$ is transverse by the Bianchi identity.  Moreover $f^{(n)bc}$ is compactly supported between $\Sigma$ and $\Sigma'$ because  $h^{(n,\chi)}_{b'c'}$ satisfies the linearized Einstein equations in the future of $\Sigma$ and vanishes in the past of $\Sigma'$.  Now,
\begin{eqnarray}
(f^{(n)},\hat{h})_{\rm st}
& = & \int \sqrt{-g}(-\hat{h}_{bc}\nabla_a p^{(n,\chi)abc}
+ S^{bcb'c'}\hat{h}_{bc}h^{(n,\chi)}_{b'c'}) \nonumber \\
& = & \int \sqrt{-g}\nabla_a(
h^{(n,\chi)}_{bc}\hat{p}^{abc} - p^{(n,\chi)abc}\hat{h}_{bc}).
\end{eqnarray}
Then by the generalized Stokes theorem we find
$(f^{(n)},\hat{h})_{\rm st} = (h^{(n)},\hat{h})_{\rm symp}$.

This observation implies that one does not lose any gauge-invariant information contained in the Wightman two-point function on a state $|\omega\rangle$,
\begin{equation}
\Delta_{aba'b'}(x,x') =  \langle\omega|\hat{h}_{ab}(x)\hat{h}_{a'b'}(x')|\omega\rangle,
\end{equation}
by considering only its smeared version, which is gauge invariant:
\begin{eqnarray}
G_g(f^{(1)},f^{(2)}) & = & \int d^Dx\sqrt{-g(x)} \int d^Dx'\sqrt{-g(x')}\nonumber \\ &&\times f^{(1)ab}(x)f^{(2)a'b'}(x')
\Delta_{aba'b'}(x,x'),\nonumber \\
 \label{graviton}
\end{eqnarray}
where the tensors $f^{(1)ab}$ and $f^{(2)ab}$ are compactly supported and transverse.

Now, for the gauge-invariant two-point function for non-interacting electromagnetic field $A_a$ analogous to Eq.~(\ref{graviton}), the two-point function $\langle\omega|A_a(x)A_{a'}(x')|\omega\rangle$ is smeared by smooth and compactly-supported transverse vectors $f^{(1)a}(x)$ and $f^{(2)a'}(x')$.  If the spacetime is topologically trivial, or contractible, then one can find smooth and compactly-supported antisymmetric tensors $u^{(1)ab}$ and $u^{(2)ab}$ such that
$f^{(1)a} = 2\nabla_b u^{(1)ab}$ and $f^{(2)a}= 2\nabla_b u^{(2)ab}$.  (It can be shown that the difference between the supports of $u^{(i)ab}$ and $f^{(i)a}$ can be made arbitrarily small.) Then, by integration by parts, the smeared two-point function is shown to be equal to the two-point function of the field-strength tensor,
$\langle\omega|F_{ab}(x)F_{a'b'}(x')|\omega\rangle$, $F_{ab}:=\nabla_a A_b - \nabla_b A_a$, smeared with $u^{(1)ab}(x)$ and $u^{(2)a'b'}(x')$~\cite{Dimock}. In the next two sections we
show that in Minkowski space and in the Poincar\'e patch of de~Sitter space the gauge-invariant graviton two-point function $G_g(f^{(1)},f^{(2)})$ is equal to the two-point function for the linearized Weyl tensor smeared with compactly-supported tensors.

\section{Equivalence in Minkowski space} \label{Minkowski}

In Minkowski space the linearized Weyl tensor, which equals the linearized Riemann tensor, is given in terms of the graviton field $\hat{h}_{ac}$ as
\begin{equation}
\hat{C}_{abcd}^{\rm flat} = \kappa (\partial_a\partial_{[c}\hat{h}_{d]b} - \partial_b \partial_{[c}\hat{h}_{d]a}), \label{Weyltensor}
\end{equation}
where $\kappa<0$ is a constant. Here we use $\partial_a$ instead of $\nabla_a$ to emphasize that
$[\partial_a,\partial_b]V^c=0$.  Now, suppose that for any smooth and compactly-supported symmetric tensor $f^{ac}$ one can find a smooth and compactly-supported tensor $v^{abcd}$ antisymmetic under $a\leftrightarrow b$ and $c\leftrightarrow d$ and symmetric under $[ab]\leftrightarrow [cd]$
such that $f^{ac} = \partial_b\partial_d v^{abcd}$.  Then by substituting
$f^{(i)ac}=\partial_b\partial_d v^{(i)abcd}$, $i=1,2$, into Eq.~(\ref{graviton}) and integrating by parts, one finds that $G_g(f^{(1)},f^{(2)})$ is expressed as the two-point function of the linearized Weyl tensor smeared with $v^{(1)abcd}$ and
$v^{(2)a'b'c'd'}$.
We show that any smooth and compactly-supported symmetric tensor $f^{ac}$ can indeed
be expressed as $f^{ac} = \partial_b\partial_d v^{abcd}$, where $v^{abcd}$ has the properties mentioned above.  (In this and next sections all tensors are smooth and compactly supported unless otherwise stated.) It is sufficient to find a tensor $u^{abc}$, antisymmetric under $a\leftrightarrow b$ and satisfying $\partial_c u^{abc} = 0$ such that $f^{ac} = \partial_b u^{abc}$.  This is because then we can write $u^{abc} = \partial_d \tilde{v}^{abcd}$ for some $\tilde{v}^{abcd}$, antisymmetric under $c\leftrightarrow d$ and that the tensor
$v^{abcd} = (\tilde{v}^{abcd} + \tilde{v}^{cdab})/2$ will be a tensor with the desired symmetries satisfying $f^{ac} = \partial_b\partial_d v^{abcd}$.

The metric signature plays no r\^ole in finding the tensor $u^{abc}$ with the properties described above.
%Now we show that $f^{ac} = \partial_b u^{abc}$ with $u^{abc}$ satisfying
%$u^{abc} = -u^{bac}$ and $\partial_c u^{abc} =0$.  Note that
%the metric signature plays no role in this problem.
We work in $D$-dimensional Euclidean space. We first solve the following equation for $\gamma_{ac}$:
\begin{equation}
\partial_b\partial_d Y^{abcd} = f^{ac}, \label{EinsteinY}
\end{equation}
where
\begin{equation}
Y_{abcd}:= - \delta_{a[c}\gamma_{d]b} + \delta_{b[c}\gamma_{d]a}
+ \delta_{a[c}\delta_{d]b}\gamma.
\end{equation}
Here, the tensor $\delta_{ab}$ is the metric on the Euclidean space and $\gamma:=\delta^{ab}\gamma_{ab}$. (The tensor $\gamma_{ab}$ is not compactly supported.)   Eq.~(\ref{EinsteinY}) is basically the linearized Euclidean Einstein equations with $f^{ac}$ as the source tensor.
We define
\begin{eqnarray}
U_{abc}  & := &  \partial^d Y_{abcd} \nonumber \\
 & = & \partial_{[a} \gamma_{b]c} - \delta_{c[a}(\partial^d \gamma_{b]d}
 - \partial_{b]}\gamma).
\end{eqnarray}
Then we have $\partial_b U^{abc} = f^{ac}$ and $\partial_c U^{abc}=0$.  Although the tensor $U^{abc}$ is not compactly supported, we show below that a compactly supported tensor $u^{abc}$ with the same symmetries and satisfying the same differential equations as $U^{abc}$ can be constructed.

We work in polar coordinates with $r^2:= \delta_{ab}x^a x^b$.
Since $f^{ab}$ is compactly supported, we have $f^{ab}(x)=0$
if $r>R$ for some $R>0$.  Then, one can choose a gauge condition such that
$\gamma_{ab}$ is transverse, traceless and with vanishing radial components, $\gamma_{ab}x^b = 0$, for $r>R$.  This result is demonstrated in the Appendix.  Then
$U_{abc} = \partial_{[a}\gamma_{b]c}$.
We define the following differential and integral operators on any (non-compactly-supported) tensor $\Phi$ (with the indices omitted)  decaying faster than $r^{-N}$ for large $r$:
\begin{eqnarray}
\mathcal{D}_N\Phi & := & -(x^d\partial_d + N)\Phi,\\
\mathcal{G}_N \Phi & : = & r^{-N}\int_r^\infty \rho^{N-1}\Phi\,d\rho, \label{GN}
\end{eqnarray}
where in Eq.~(\ref{GN}) $r$ is replaced by $\rho$ in the integrand.  Then
$\mathcal{D}_N\mathcal{G}_N\Phi = \mathcal{G}_N \mathcal{D}_N\Phi = \Phi$.  We also find
\begin{equation}
\partial^a(\mathcal{G}_N A_a) = \mathcal{G}_{N+1}\partial^a A_a
\end{equation}
if $A_a$ decays faster than $r^{-N}$ for large $r$.

%We expand $\gamma_{ab}$ in power series in $r^{-1}$:
%%\begin{equation}
%$\gamma_{ab} = \sum_{\ell=2}^\infty \gamma_{ab}^{(\ell)}$,
%%\end{equation}
%where $\gamma_{ab}^{(\ell)}$ is of order $r^{-(\ell+D-2)}$, i.e.
%%\begin{equation}
%$x^c\partial_c \gamma_{ab}^{(\ell)} = -(\ell+D-2)
%\gamma_{ab}^{(\ell)}$.
%%\end{equation}
%(The series starts from $\ell=2$ though this fact is not important in our argument.)  Thus,
%%\begin{equation}
%$U_{abc}  =
%\sum_{\ell=2}^\infty U^{(\ell)}_{abc}$,
%%\end{equation}
%where $2U^{(\ell)}_{abc}:= \partial_a \gamma_{bc}^{(\ell)} - \partial_b \gamma_{ac}^{(\ell)}$.
%One can readily show that
%$U^{(\ell)}_{abc}x^c = 0$.

Now we define for $r> R$
\begin{equation}
M_{dabc}: = \mathcal{G}_{D-3}(x_d U_{abc} + x_a U_{bdc} + x_b U_{dac}).
\end{equation}
The operator $\mathcal{G}_{D-3}$ is well defined because the tensor $\gamma_{ab}$ decays like
$r^{-D}$ or faster for large $r$.
%:= \sum_{\ell=2}^\infty
%\frac{1}{\ell+1}(x_d U_{abc}^{(\ell)}
%+ x_a U_{bdc}^{(\ell)}
%+ x_b U_{dac}^{(\ell)}).
%\end{equation}
One can readily show that
$\partial^d M_{dabc} = -U_{abc}$.  Let
$\chi(r)$ be a smooth function defined for $r>R$ such that $\chi(r)=1$
for $r> R'> R$ and $\chi(r)=0$ for $R'>  R''> r >R$.  Let
\begin{equation}
u_{abc}
:= U_{abc} + \partial^d(\chi M_{dabc}).
\end{equation}
Then $u_{abc} = 0$ for $r>R'$, i.e.\ $u_{abc}$ is compactly supported. Since $M_{dabc}$ is antisymmetric in the indices $d$, $a$ and $b$, we have
$\partial_b u^{abc} = f^{ac}$. Moreover, $\partial_c u^{abc}=0$ because
$\partial_c U^{abc}=0$ and $\partial^c M_{dabc} = M_{dabc} x^c = 0$. Thus, we have constructed a tensor
$u^{abc}$ that satisfies all the properties necessary to show that the gauge-invariant graviton two-point function is equivalent to the linearized Weyl-tensor two-point function as stated at the start of this section.

\section{Equivalence in de~Sitter space} \label{deSitter}

The metric of the Poincar\'e patch of de~Sitter space is
\begin{equation}
ds^2 = \Omega^2(-d\eta^2 + d\mathbf{x}^2),\,\,\, \eta \in (-\infty,0),
\end{equation}
where $\Omega = (H|\eta|)^{-1}$.  In this spacetime it can readily be shown that, if
\begin{equation}
f^{ac} = \nabla_b \nabla_d V^{abcd} + H^2 g_{bd}V^{abcd},  \label{aiming}
\end{equation}
where
\begin{equation}
V^{abcd} = V^{[ab][cd]} = V^{[cd][ab]},  \label{V-symmetry}
\end{equation}
then $f^{ac}$ is transverse.  In this section we show that if $f^{ac}$ is smooth, compactly-supported in the Poincar\'e patch and transverse, then there is a smooth and compactly-supported tensor $V^{abcd}$ with the symmetry (\ref{V-symmetry}) such that Eq.~(\ref{aiming}) is satisfied.  This will imply that, if the linearized gravity operator $\hat{h}_{ab}$ satisfies the linearized Einstein equations, $E^{(1)}_{ac}(\hat{h}) = 0$, where
\begin{eqnarray}
E_{ac}^{(1)}(\hat{h}) & = & H^2 \hat{h}_{ac} + \tfrac{D-3}{2}H^2g_{ac} \hat{h} \nonumber \\
&& + \tfrac{1}{2}\left( \nabla_a \nabla^b  \hat{h}_{bc}+ \nabla_c \nabla^b \hat{h}_{ab} - \nabla_a\nabla_c \hat{h}
\right. \nonumber \\
&& \left. - \Box \hat{h}_{ac}
- g_{ac}\nabla^b \nabla^d \hat{h}_{bd} + g_{ac}\Box \hat{h}\right), \label{Einstein}
\end{eqnarray}
then
\begin{eqnarray}
\int d^Dx\sqrt{-g}\,f^{ac}\hat{h}_{ac} & = & \int d^Dx \sqrt{-g}\, V^{abcd}(\nabla_b\nabla_d \hat{h}_{ac} + g_{bd}\hat{h}_{ac}) \nonumber \\
& = & \kappa^{-1}\int d^Dx\sqrt{-g}\,V^{abcd}C_{abcd}(\hat{h}), \label{Weyl-reduction}
\end{eqnarray}
where $C_{abcd}(\hat{h})$ is the linearized Weyl tensor in the de~Sitter background.

First we show that any symmetric tensor $f^{ac}$ in de~Sitter space can be written as
\begin{equation}
f_{ac} = f^{(\textrm{tl})}_{ac} + E^{(1)}_{ac}(\gamma^{(t)}), \label{decompose}
\end{equation}
where the tensor $\gamma^{(t)}_{ac}$ and the transverse-traceless
tensor $f^{(\textrm{tl})}_{ac}$ are symmetric.  To see this we let
\begin{equation}
\gamma^{(t)}_{ab}:= \tfrac{2}{(D-1)(D-2)H^2}\left[(D-1)f_{ab}-g_{ab}f\right],
\end{equation}
where $f:=g_{ac}f^{ac}$.  Then
\begin{eqnarray}
E^{(1)}_{ac}(\gamma^{(t)}) & = &  \tfrac{1}{(D-2)H^2}\left[\Box f_{ac} - \tfrac{1}{D-1}(g_{ac}\Box f-\nabla_a\nabla_c f)\right]\nonumber\\
&& -\tfrac{1}{D-2}\left(2f_{ac}-g_{ac}f\right).
\end{eqnarray}
It can readily be verified that $f^{(\textrm{tl})}_{ac} = f_{ac} - E^{(1)}_{ac}(\gamma^{(t)})$ is traceless.  Note also
\begin{equation}
\int d^Dx\sqrt{-g}\,f^{ac}\hat{h}_{ac} = \int d^Dx \sqrt{-g}\,f^{(\textrm{tl})ac}\hat{h}_{ac}
\end{equation}
because
\begin{eqnarray}
\int d^Dx \sqrt{-g}\,E^{(1)ac}(\gamma^{(t)})\hat{h}_{ac}
& = & \int d^Dx\sqrt{-g}\,\gamma^{(t)ac}E^{(1)}_{ac}(\hat{h})\nonumber \\
&  = & 0.
\end{eqnarray}

Next we observe that for any symmetric tensor $\gamma_{ac}$ we have
\begin{equation}
E^{(1)}_{ac}(\gamma) = \nabla^b \nabla^d \Gamma_{abcd} + H^2 g^{bd}\Gamma_{abcd},  \label{ein-trivial}
\end{equation}
where
\begin{equation}
\Gamma_{abcd} := -g_{a[c}\gamma_{d]b} + g_{b[c}\gamma_{d]a} + (g_{a[c}g_{d]b} - g_{b[c}g_{d]a})\gamma \label{Gamma-def}
\end{equation}
with $\gamma := g^{ac}\gamma_{ac}$. Using Eq.~(\ref{ein-trivial}) in Eq.~(\ref{decompose}), we find
\begin{equation}
f_{ac} =  f^{(\textrm{tl})}_{ac} + \nabla_b\nabla_d \Gamma^{(t)abcd} + H^2 g_{bd}\Gamma^{(t)abcd},
\end{equation}
where $\Gamma^{(t)}_{abcd}$ is given by Eq.~(\ref{Gamma-def}) with $\gamma_{ac} = \gamma^{(t)}_{ac}$.
Thus, if
\begin{equation}
f_{ac}^{(\textrm{tl})} = \nabla_b\nabla_d V^{(\textrm{tl})abcd}  \label{aim2}
\end{equation}
with a traceless tensor $V^{(\textrm{tl})abcd}$ with the symmetry (\ref{V-symmetry}),
then Eq.~(\ref{aiming}) will be satisfied with
\begin{equation}
V_{abcd} = V^{(\textrm{tl})}_{abcd} + \Gamma^{(t)}_{abcd}.
\end{equation}
We show below that Eq.~(\ref{aim2}) holds.

First we note that the equation $\nabla_a f^{(\textrm{tl})ac} = 0$ can be written
$\partial_a (\Omega^{D+2}f^{(\textrm{tl})ac}) = 0$ in any conformally-flat spacetime for a symmetric traceless tensor $f^{(\textrm{tl})ac}$~\cite{WaldGR}.
By our result in Minkowski space in the previous section this implies that we can write
\begin{equation}
\Omega^{D+2}f^{(\textrm{tl})ac} = \partial_b\partial_d v^{abcd}, \label{v-equation}
\end{equation}
where the tensor $v^{abcd}$ has the symmetry
$v^{abcd} = v^{[ab][cd]} = v^{[cd][ab]}$.
Denoting the traceless part of $v^{abcd}$ by $v^{({\rm tl})abcd}$ and defining
$v^{ac} := \eta_{bd}v^{abcd}$ and $v:=\eta_{ac}v^{ac}$, where $\eta_{ac}$ is the standard metric tensor on Minkowski space,
we find
\begin{eqnarray}
v^{abcd} & = & v^{({\rm tl})abcd} + \tfrac{2}{D-2}(\eta^{a[c}v^{d]b} - \eta^{b[c}v^{d]a})\nonumber \\
&& - \tfrac{1}{(D-2)(D-1)}(\eta^{ac}\eta^{bd}-\eta^{ad}\eta^{bc})v.  \label{expansion}
\end{eqnarray}
Substitution of this expression into Eq.~(\ref{v-equation}) yields
\begin{equation}
\Omega^{D+2}f^{(\textrm{tl})ac} = \partial_b\partial_d v^{(\textrm{tl})abcd} + \tfrac{1}{D-2}s^{ac},
\end{equation}
where
\begin{equation}
s^{ac}  :=  \Box v^{ac} - \partial^a \partial_b v^{cb}
- \partial^c \partial_b v^{ab}
+ \tfrac{1}{D-1}(\partial^a\partial^c v - \eta^{ac}\Box v).
\end{equation}

Now, since $\partial_b\partial_d v^{abcd} = \Omega^{D+2}f^{(\textrm{tl})ac}$ is traceless, the tensor $v^{ac} = \eta_{bd}v^{abcd}$ satisfies
$\partial_a\partial_c v^{ac}=0$.  That is, $\partial_c v^{ac}$ is transverse.  This means that
$\partial_c v^{ac} = \partial_c q^{ac}$, i.e.\ $\partial_c (v^{ac}-q^{ac})=0$,
where $q^{ac}$ is an antisymmetric tensor. This in turn implies that
%\begin{equation}
$v^{ac} - q^{ac} = 2\partial_b w^{abc}$, %\label{ant}
%\end{equation}
where $w^{abc} = w^{a[bc]}$. By symmetrizing %Eq.~(\ref{ant})
this equation we obtain
\begin{equation}
v^{ac} = \partial_b w^{abc} + \partial_b w^{cba}.  \label{defw}
\end{equation}
We define
\begin{equation}
W^{abcd}:= \partial^c w^{dab}-\partial^d w^{cab} + \partial^a w^{bcd} - \partial^b w^{acd}.
\end{equation}
Then
\begin{eqnarray}
\eta_{bd}W^{abcd} & = & \eta_{bd}(\partial^c w^{dab} + \partial^a w^{dcb}) + v^{ac},\\
\eta_{ac}\eta_{bd}W^{abcd} & = & 2v.
\end{eqnarray}
Let $W^{(\textrm{tl})abcd}$ be the traceless part of $W^{abcd}$. Then, after some algebra we find
\begin{equation}
\partial_b \partial_d W^{(\textrm{tl})abcd} = \tfrac{D-3}{D-2}s^{ac}.
\end{equation}
Hence by defining
\begin{equation}
\check{v}^{(\textrm{tl})abcd} : = v^{(\textrm{tl})abcd} + \tfrac{1}{D-3}W^{(\textrm{tl})abcd}
\end{equation}
we have
\begin{equation}
\Omega^{D+2}f^{(\textrm{tl})ac} = \partial_b\partial_d \check{v}^{(\textrm{tl})abcd},
\end{equation}
where $\check{v}^{(\textrm{tl})}_{abcd} = \check{v}^{(\textrm{tl})}_{[ab][cd]} = \check{v}^{(\textrm{tl})}_{[cd][ab]}$.
Now, it can readily be shown that
\begin{equation}
\partial_b \partial_d \check{v}^{(\textrm{tl})abcd} = \Omega^{D+2}\nabla_b\nabla_d (\Omega^{-D-2}\check{v}^{(\textrm{tl})abcd})
\end{equation}
using $\Omega = (H|\eta|)^{-1}$ and the fact that $\check{v}^{(\textrm{tl})}_{abcd}$ is traceless.  Hence Eq.~(\ref{aim2}) is satisfied with
\begin{equation}
V^{(\textrm{tl})abcd} = \Omega^{-D-2}\check{v}^{(\textrm{tl})abcd}.
\end{equation}
This establishes Eq.~(\ref{aiming}). Then Eq.~(\ref{Weyl-reduction}) implies that
the gauge-invariant graviton two-point function in de~Sitter space
can be expressed as
\begin{eqnarray}
G_g(f^{(1)},f^{(2)})
& = & \kappa^{-2}\int d^D x \sqrt{-g(x)}\int d^D x' \sqrt{-g(x')} \nonumber \\
&& \times V^{(1)abcd}(x)V^{(2)a'b'c'd'}(x')\nonumber \\
&& \times \langle \omega|\hat{C}_{abcd}(\hat{h}(x))\hat{C}_{a'b'c'd'}(\hat{h}(x'))|\omega\rangle \nonumber \\
\end{eqnarray}
if each of $f^{(1)}_{ac}$ and $f^{(2)}_{ac}$ has support in a Poincar\'e patch.
This result was used recently in proving a ``cosmic no hair theorem" for linearized quantum gravity~\cite{Morrison}.
From our construction of $V^{abcd}$ from $f^{ab}$, it is clear that
the difference between the supports of $V^{abcd}$ and $f^{ab}$ can be made arbitrarily small.
%if the support of $f^{ab}$ is within an open set, then we can construct%
%$\check{v}^{abcd}$ in such a way that its support is also within this open set.

\section{Discussion} \label{discuss}

In this paper we showed that the gauge-invariant graviton two-point function of Ref.~\cite{FewsterHunt} is equivalent to the two-point function of the linearized Weyl tensor in Minkowski space and in the Poincar\'e patch of de~Sitter space.  This result is analogous to the following result for the non-interacting electromagnetic field: the gauge-invariant photon two-point function is equivalent to the two-point function of the field-strength tensor in topologically trivial, or contractible, spacetime.

Recently the Weyl-tensor two-point function in the Poincar\'e patch of de~Sitter space has been computed by Mora and Woodard~\cite{Mora:2012kr}.  Their result agrees with that computed in the
covariant gauge by Kouris~\cite{Kouris} (with corrections to be published)~\cite{private}.  This two-point function between points $(\eta,\mathbf{x})$ and $(\eta',\mathbf{x}')$  decays like $H^2\|\mathbf{x}-\mathbf{x}'\|^{-4}$ as $\|\mathbf{x}-\mathbf{x}'\|$ tends to infinity.
%and like $|\eta-\eta'|^{-4}$ as
%$|\eta-\eta'|\to \infty$.
Since this two-point function is equivalent to the gauge-invariant
graviton two-point function as we have shown in this paper, the latter decays in the same way for large $\|\mathbf{x}-\mathbf{x}'\|$.
%It has been confirmed recently that the covariant graviton two-point function and the two-point function in the transverse-traceless-synchronous gauge in global coordinates give the same two-point function for the linearized Weyl tensor~\cite{FaizalHiguchi}.  Assuming that the graviton two-point function in the transverse-traceless-synchronous gauge in the Poincar\'e patch is also equivalent to the covariant two-point function, we are led to the conclusion
%the gauge-invariant graviton two-point function has the same decay properties as the two-point function for the linearized Weyl tensor.
It will be interesting to investigate whether our result is compatible with the claim that the logarithmic growth of the graviton two-point function leads to physical consequences such as instability of de~Sitter space. (For recent works on this subject, see, e.g., Refs.~\cite{GiddingsSloth1,GiddingsSloth2,GiddingsSloth3}).

\acknowledgments
%%%%%%%%%%%%%%%%%%%%%%%%%%%%%%%%%%%%%%%%%%%%%%%%%%%%%%%%%%%%%%%

We thank Dave Hunt, Don Marolf, Ian Morrison and Richard Woodard for useful discussions.

\appendix

\section*{Appendix: Gauge fixing in linearized Euclidean gravity} %\label{AppendixA}

In this Appendix we prove that the gauge conditions $x^a \gamma_{ab} = 0$ can be imposed as well as the transverse-traceless conditions on the solutions to linearized Euclidean Einstein equations in the vacuum.

First we impose the de~Donder gauge condition (which can be achieved by solving a Poisson equation).
%Then the trace $\gamma := {\gamma^a}_a$ satisfies $\Box\gamma=0$.
The tensor $\overline{\gamma}_{ab} = \gamma_{ab}-\tfrac{1}{2}\delta_{ab}\gamma$ satisfies
$\Box \overline{\gamma}_{ab}=0$.  Hence $\Box \gamma = 0$.  There are no monopole solutions
$\overline{\gamma}_{ab}=K_{ab}\psi(r)$, where $K_{ab}$ is a constant tensor, to $\Box \overline{\gamma}_{ab} = 0$ satisfying the de~Donder condition.  This implies that $\gamma_{ab}$ decays at least as fast as $r^{-(D-1)}$ for large $r$.

Define
\begin{equation}
B_a := r^2\partial_a\gamma-2x_ax^b\partial_b\gamma-(D-2)x_a\gamma.
\end{equation}
This vector decays at least as fast as $r^{-(D-2)}$ for large $r$.
Then we find $\Box B_a = 0$ and
$\partial_a B^a = - 2\mathcal{D}_{D/2}\mathcal{D}_{D-2}\gamma$.  Hence, if we let
\begin{equation}
A_a : = - \mathcal{G}_{D-3}\mathcal{G}_{(D-2)/2}B_a,
\end{equation}
which is well defined,
then $\gamma^{(g)}_{ab}:= \frac{1}{4}(\partial_a A_b + \partial_b A_a)$ satisfies the de~Donder condition and
the equation ${\gamma^{(g)a}}_a = \gamma$.  Thus, the trace can be gauged away.
%%
%
%  Hence, it is sufficient to show that the condition $x^a \gamma_{ab} = 0$ can be imposed on the solutions of the %form
%%\begin{equation}
%$\gamma^{(\ell)}_{ab} = r^{-(2\ell+D-2)}k_{ab:c_1\cdots c_\ell}x^{c_1}\cdots x^{c_\ell}$,
%%\end{equation}
%where $\delta^{c_1c_2}k_{ab:c_1c_2\cdots c_{\ell}} = 0$~\cite{ChodosMyers,RubinOrdonez}.
%
%It is clear that the de~Donder condition cannot be satisfied if $\ell=0$. Thus, we can assume $\ell\geq 1$.
%Recall that for a gauge solution, $\gamma_{ab}^{(g)} = \partial_a A_b + \partial_b A_a$ the de~Donder gauge %condition reduces to $\Box A_a = 0$.  This equation is satisfied by
%\begin{equation}
%A^{(\ell)}_{a} = -\frac{1}{2(2\ell+D-4)r^{2\ell+D-4}}\delta^{mn}k_{mn:ac_1\cdots c_{\ell-1}}x^{c_1}\cdots %x^{c_{\ell-1}}.
%\end{equation}
%Moreover, for this solution $2\partial^a A_a^{(\ell)} = \delta^{ab}\gamma^{(\ell)}_{ab}$.  Thus, the trace can be %gauged away.
That is, $\gamma_{ab}$ can now be assumed to be transverse and traceless.  Then, if we define
\begin{equation}
\tilde{\gamma}_{ab}
:= \gamma_{ab} - \partial_a\partial_b (x^cx^d\mathcal{G}_1\mathcal{G}_2\gamma_{cd}),
\end{equation}
then $x^ax^b \tilde{\gamma}_{ab} = 0$ and $\tilde{\gamma}_{ab}$ is transverse and traceless. Finally we define
\begin{equation}
\check{\gamma}_{ab}
:= \tilde{\gamma}_{ab} +
\left[\partial_a(x^c\mathcal{G}_0\tilde{\gamma}_{bc}) + \partial_b(x^c\mathcal{G}_0\tilde{\gamma}_{ac})\right].
\end{equation}
Then we find $x^b\check{\gamma}_{ab} = 0$, $\partial^b \check{\gamma}_{ab} = 0$ and
$\delta^{ab}\check{\gamma}_{ab}=0$.

\end{document}